# Remote which-way-choice scheme: proposed foundations experiments

Artur Szczepański

Institute of Fundamental Technological Research,
Polish Academy of Sciences,
Świętokrzyska 21, 00-049 Warszawa, Poland.
E-mail: aszczep@ippt.gov.pl

**Abstract.** Twin photons from spontaneous parametric down-conversion with preselected polarization are used as spatially disjoint subsystems. One photon is subject to an interference measurement in a Mach-Zehnder interferometer, which does not decohere the system, while a projective measurement in the polarization space of the second twin decoheres the system, and induces the which-way choice of the first photon. This would be an experimental example of way-choice free of direct action upon the interfering photon. Two further applications of the proposed scheme are considered: a new method of quantum information speed measurement, and a test of commutation of space-like remote measurements.

## 1. Introduction

Since the early days of quantum mechanics the double-slit experiment has been at the center of foundations questions. The Bohr-Einstein debate a well-known historical example [1]. Mature quantum theory provides the famous Feynman's remark on the "mystery of quantum mechanics" contained in this experiment [2]. In the last decades of the twentieth century matter waves interferometry have opened experimental insight into new foundations issues (see, e.g. [3]). In particular, one more aspect of this "quantum mystery" has been seen in neutron interference experiments: the parts of a single split neutron subject to the spin flipping electromagnetic interaction behaved simultaneously as undivided single neutrons[(1)] [4]. Recent technological achievements shifted a

---

[(1)] *(In the present text the footnotes are inserted in the main text, which is convenient in case of screen-reading.)* The latter remark may be taken for a rather trivial one: actually, in any interference experiment, in which the split beams are deflected toward the recombination region to form an interference pattern, the interaction deflecting the *partial* wave packets of a single quanton is always formally described by standard quantum mechanics as an action upon *undivided quantons*. In the mentioned neutron case, the flipper's field is under active control. Its formal action on the undivided magnetic moment of the neutron is explicit in the calculation of the predicted (and seen) shape of the interference pattern. This makes it more persuasive.

---

number of foundation issues from the earlier *gedanken* level to real experiments. We have witnessed new electron interference experiments, atom and fullerene interferences, and experimental tests of decoherence via fringes visibility loss by means of engineered decoherence action [5-10].

Quantum optics has recently provided a new, powerful tool for foundations research: experimentally controlled entangled systems (for an overview, see, e.g. [11], and [12] for an instructive account of applications to EPR-Bell tests). Appling the latter to the double-slit which-way scheme opens access to new experimental evidence. Accordingly, I propose here a novel type of the which-way experiment. It consists in using entangled photon pairs, one photon of which is subject to a conventional Mach-Zehnder interference measurement (the double-slit experiment), while a projective, decohering action upon the second correlated photon forces the first photon to make the which-way choice modifying the observed interference pattern. In contrast to the existing experiments and *gedanken* arguments, the proposed which-way choice is made without any direct action upon (or of) the interfering photon. This would provide a new experimental argument in the recent debate, in which the question of the momentum transfer washing out the interference fringes has been a central one, and would shed light on the relation of complementarity to the uncertainty relation [13-16]. While I regard the latter as the most direct use of the proposed scheme, a number of other applications to foundations tests seems to be of major interest, as well. Two examples are pointed out here: a new method of quantum information speed measurement (see, [17] for the existing results), and a test of commutation of spatially disjoint measurements. The latter would make sense if one of the involved measurements were non-demolishing in the state space of the system, in which the second measurement were a usual projective one. This remarkable feature is realized in the discussed here remote which-way choice scheme: the projective measurement of the polarization state of one photon decoheres the entangled biphoton polarization state, while the interference measurement with the second twin photon in the Mach-Zehnder interferometer does not decohere the system's polarization state. I shall refer to this kind of non-demolishing measurement to as coherence conserving measurement.

The organization of the paper: In Sec. 2 the proposed scheme, and its expected basic result, that is the remote which-way choice, is described. To realize the latter "asymptotic actions" of Alice (Alice is the party, who makes the which-way-choice-generating projective measurement; Bob operates the interferometer) are expected to be sufficient: Alice's projective measurement follows immediately the preparation stage at the source, which may be referred to as "0-time", or Alice's detectors are shifted to infinity (removed), thus the "asymptotic ∞-time" of Alice's action.

In the application of the scheme to the further proposed foundations experiments the "asymptotic" conditions must be abandoned, and replaced by Alice's actions shifted from "zero" and "infinity" to distances, that is time, comparable with that of Bob's measurement. Specifically, the experimental determination of the time interval between Alice's projective measurement and the emergence time of its effect seen in Bob's interference measurement is necessary. A procedure to do this is outlined in Sec. 3. A new method of the measurement of the quantum information speed is proposed as a direct application of this result (Sec. 4). A direct way to test the non-decohering character in the polarization state space of Bob's interference measurement is discussed in Sec. 5. Its expected positive result is the basis for understanding the predicted result of the commutation test of Alice's projective measurement (in the polarization space) and Bob's coherence-conserving interference measurement (Sec. 6). Some conclusions are outlined in Sec. 7.

## 2. Which-way choice scheme via polarization entanglement.

In the proposed here experiment a bipartite entangled quantum system consisting of twin photons from spontaneous parametric down-conversion is used to realize the remote way-choice. One photon, say the signal photon, is subject to a projective measurement, which decoheres the total system inducing the which-way choice of the idler photon split in a Mach-Zehnder interferometer. The scheme of the experimental setup is shown in Fig. 1.

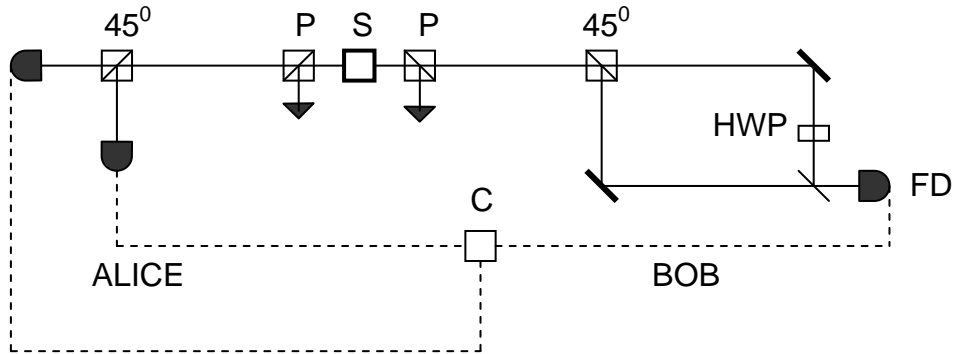

Fig. 1. Scheme of the remote which-way choice setup. S is a source of twin photons (e.g. from type-II spontaneous parametric down-conversion) with preset by P polarization planes: the polarizing beam splitters P, which polarization planes of the transmitted beams are set mutually perpendicular, define the preset polarization plane of the photon propagating toward Alice, and that propagating toward Bob. The reflected beams are absorbed at P's output ports. Alice's and Bob's polarizing beam splitter are rotated $45^0$ relative to the polarization plane of the incoming photons. Alice's split photons are detected behind the beam splitter's output ports. Bob's split photons are recombined by the second beam splitter of the Mach-Zehnder interferometer. HWP is a half wave plates system rotating the polarization plane of a partial beam (here, the upper), so that the polarization planes of the recombined beams are parallel. FD is the detector seeing interference fringes. Alice's and Bob's detectors are linked by coincidence circuits C.

S is the source of twin photons, which are directed in separate spatial modes. The polarizing beam splitters P at the output ports of S are set accordingly to the maximal polarization correlation of the twins: the polarization planes of the transmitted beams are set parallel if S is a type-I phase matching source, or set mutually perpendicular for a type-II source (see, e.g. [18]). Both the reflected photons are absorbed at the output ports of P. Alice (left-hand-wing side, spatial A-mode) operates the polarizing beam splitter, which is rotated $45^0$ relative to the polarization of the incoming beam, and operates the detectors behind the beam splitter output ports. Bob operates the Mach-Zehnder interferometer and the interference-pattern-seeing detector FD in the right-hand-wing side (spatial B-mode). The interferometer's input polarizing beam splitter is rotated $45^0$ relative to the incoming beam polarization plane. A system of half wave plates rotates the polarization plane in one of the interferometer's partial beams so that the recombined beams are parallel polarized. The action of the elements in both the A-setup and the B-setup (apart from detectors, of course) is assumed to be unitary. Alice's detectors are linked to Bob's detector by (classical) coincidence circuits.

The setup can be regarded as operating in three steps: (i) the preparation, (ii) unitary operations in the A-setup and in the B-setup, (iii) projective intervention (measurement) in the A-subsystem, and interference fringes observation in the B-subsystem.

*(i) The preparation.*

The action of the beam splitters P and the (projective) action of the absorbers placed at P's reflection output ports prepare the initial polarization state of the photon pair propagating along the A-mode optical path, and along the B-path as:

$$|initial\rangle = |A;\varphi\rangle |B;\varphi+\theta\rangle. \qquad (1)$$

$\varphi$ is the angle of the A-mode-photon polarization plane (set by P) relative to the chosen reference frame. $\theta = 0$, if S is a type-I source; $\theta = 90^0$, if S is a type-II source. The B-mode-photon is then parallel polarized relative to the A-photon, for a type-I S; the A-photon and B-photon are mutually perpendicular polarized when S is of type-II, which is the usual relation of polarization planes of correlated twin photons. Here, however, this relation is more specific: the orientation (relative to the setup) of these planes is preset by the action of P.

Note by the way, that so prepared states are low noise states, since the conditional probability of finding the B-twin in the B-mode, if the A-photon were found in the A-mode, and the probability of finding the B-twin in the B-mode, if the A-photon were reflected by P, satisfy the relation:

$$P(B_{PT} | A_{PT}) \gg P(B_{PT} | A_{PR}), \qquad (2)$$

where the subscripts PT and PR stand for "transmitted by P" and "reflected by P", respectively. This obviously implies that the number of single photons from S entering the final B-setup and contributing to noise in is small relative to the number of pairs contributing to the signal.

*(ii) Unitary operations in the A-setup and in the B-setup.*

Choose $\varphi$ to make an angle of $45^0$ with the plane referred to as the horizontal polarization plane. The A-photon is split, roughly 50/50, by the A-beam-splitter into a horizontally polarized (transmitted), *AH,* and a vertically polarized (reflected), *AV*, part (see, e.g. [19]):

$$|AH\rangle = \tau e^{iTA} |A;\varphi-45^0\rangle, \qquad (4)$$

$$|AV\rangle = \rho e^{iRA} |A;\varphi+45^0\rangle. \qquad (5)$$

The irrelevant here vacuum contributions have been omitted. So acts Bob's input beam splitter, and consequently the polarization state of the total system can be written as:

$$|split\rangle_I = \alpha e^{i(TA+TB)} |AH\rangle|BH\rangle + \beta e^{i(RA+RB)} |AV\rangle|BV\rangle, \quad \text{type-I S,} \quad (6)$$

or

$$|split\rangle_{II} = \gamma e^{i(TA+RB)} |AH\rangle|BV\rangle + \delta e^{i(RA+TB)} |AV\rangle|BH\rangle, \quad \text{type-II S.} \quad (7)$$

The phases *TA, TB, RA, RB,* and the constants α, β, γ, δ, are characteristic of the beam splitters (*T* for "transmitted", *A* for "A-photon", etc.). We shall refer to such states to as "reduced Bell-states", since in contrast to the usual, "non-reduced" Bell states the polarization planes are preset by the action of the beam splitters P followed by the projective absorption of the reflected beams.

Before reaching the recombining beam splitter the polarization planes of the upper and lower partial beam in the Mach-Zehnder interferometer are set parallel: a system of half waves plates HWP rotates unitarily the polarization plane of the upper (lower) partial beam. In terms of the parallel polarized upper part, $|BU\rangle$, and lower part, $|BL\rangle$, of the split B-photon state, the detection probability of the B-photon by FD can be written as,

$$P(FD) = \zeta + \eta(\langle BU|BU\rangle + \langle BL|BL\rangle + 2\sqrt{\langle BU|BL\rangle\langle BL|BU\rangle}\cos\Theta). \quad (8)$$

In the ideal case the parameter $\zeta = 0$, and a maximal, 100% visibility for the central fringes of the interference pattern would result with 50/50 beam splitters. The fringes distribution parameter, $\Theta = k\Delta r + \theta$. *k* is the wave vector of the photon's wave packet center. *Δr* is the interferometer's path difference seen by FD. $\zeta, \eta$ and $\theta$ are constants characteristic of the used optics elements (beam splitters, etc.).

*(iii) Alice's projective polarization measurement and Bob's interference fringes observation.*

Suppose first that Alice has removed her detectors. The system evolves unitarily before Bob's measurement, and the FD counts distribution approaches (with growing statistics) *P(FD)* as given by Eq. (8).

Now, if one of Alice's detectors is active, say the transmitted-beam detector, ATD, and clicks before[2] Bob's FD, then the term with $|AV\rangle$, that is

---

[2] See, [20] for a discussion concerning the assumptions behind the different meaning of "before". In this section the term "before" is understood asymptotically to ensure the time-like connection of the involved events.

---

the term containing the reflected part of the A-photon vanishes in the superposition (6) or (7). The surviving terms result in a modified (relative to (8)) conditional photocount probability of FD, measured as a coincidence rate of ATD and FD:

$$P(FD|ATD)_I = \zeta + \eta\langle BU|BU\rangle, \quad \text{type-I S,} \quad (9)$$

or

$$P(FD \mid ATD)_{II} = \zeta + \eta \langle BL \mid BL \rangle, \qquad \text{type-II S.} \qquad (10)$$

In other words, the photocount in the AT-beam projects the polarization state of the A-photon in the horizontal component, and the remote state of the B-photon in the correlated with the latter polarization component (the horizontal for type-I S, the vertical for type-II S). Consequently one of the partial wave packets in the interferometer vanishes entailing the vanishing of the interference term in the recombined wave. Thus there is no fringes in the interference pattern, or, say, FD sees an interference pattern with washed out fringes. This is interpreted here as resulting from the which-way choice of Alice's measurement.

The remote which-way choice is achieved under asymptotic conditions: Alice's A-detectors acting immediately at the source (choice made), and, on the other hand, the A-detectors removed (no choice), that is, shifted to infinity. In the next section I shall discuss an "upgrading" of the mere asymptotic conditions: the experimental specification of the time difference between the A-detector action and its earliest noticeable effect in the B-interference pattern, that is the specification of the term "before" relating Alice and Bob's measurement.

### 3. The time order of Alice's action and of its effect on Bob's measurement

As long as Alice's detectors are removed Bob sees the interference pattern as predicted by (8), since the system is evolving unitarily before Bob's measurement. Alice's detecting action before Bob's measurement is expected to decohere the polarization state of the system. In the decohered state one of the components of the split Bob's photon survives only, which should result in an observed no-fringes-pattern according to (9) and (10).

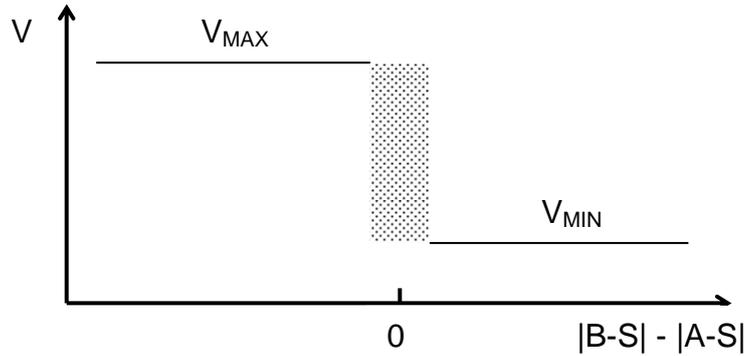

Fig. 2. The expected visibility, V, of central interference fringes as measured by Bob. |B-S|, |A-S| are optical path length from source to Bob's detector and Alice's detector, respectively. The maximal visibility, $V_{MAX}$, is seen by Bob's detector acting *before* Alice's projective measurement, $V_{MIN}$ is seen *after* Alice's measurement. In the shaded region one should expect $V_{MAX} > V > V_{MIN}$. Its width may result from a spread of the travel time of the photons from source to detectors. An ideal experiment would yield $V_{MAX} = 1$, $V_{MIN} = 0$, and a thin shaded region.

The time difference between Alice's detection and Bob's measurement can be experimentally controlled by varying the optical path length from the source to Alice's and to Bob's setup, and can be measured by the coincidence system linking Alice's and Bob's detectors. Thus one should be able to determine experimentally the meaning of the "before" of Alice's projective measurement relative to Bob's interference measurement. According to the Geneva experiments (see, [20] and refs. therein) Alice's measurement should decohere Bob's subsystem (quasi-?) instantaneously. This would result in the measured by Bob fringes visibility as shown in Fig.2.

The experiments proposed in the present paper do not essentially depend on the details of the V behavior: any kink-shaped visibility in which $V_{MAX}$ can be experimentally discriminated from $V_{MIN}$ would do. This aspect of the experimental meaning of "before" and "after" is the basis of quantum information speed measurement as discussed in the next section.

Note, however, that the visibility kink shape could yield information about the decohering process, if the Alice's acting detector were replaced by an experimentally controlled decohering environment.

### 4. Quantum information speed

The general idea of the experiment [21] discussed in this section has been inspired by the Kwiat and Chiao experiment [22]. In the present context it is a straightforward extension of the "before" and "after" experimental evaluation proposed in the previous section.

The quantum information speed, $v_{QI}$, is understood here according to the interpretation of the Gisin group [23]. There is a remarkable difference, however, in the method itself. The Geneva experiments have been based on a Franson-type correlation experiment scheme in which the measurements in both the correlated subsystems are cause-effect quasi-symmetrical: each of them can be taken for the cause-seeing-measurement, or for the effect-triggering-measurement. In particular, when both the measurements are experimentally simultaneous, they are experimentally cause-effect equivalent. On the contrary, in the proposed here experiment Alice's projective measurement is always the cause[3] of the effect observed by Bob. It is then natural to consider a quantity of

---

[3] The term "cause" is used here in a rather loose sense as the event or the short process triggering or generating a process resulting in the observed "effect". As has been already mentioned, Alice's measurement decoheres the system, Bob's measurement does not decohere the system's state, but detects Alice's decoherence effect. In this sense Alice's and Bob's actions are "cause-effect" related. The cause-effect quasi-symmetry of the Geneva experiments, and more generally, of conventional EPR-Bell tests is thus broken.

---

velocity dimension defined as the quotient of the spatial distance between Alice and Bob's measurement positions, $|A - B|$, and the time interval, $\Delta\tau$, between the "cause event" at $(\tau_A, \mathbf{A})$ and the "effect event" at $(\tau_B, \mathbf{B})$:

$$v_{QI} = \frac{|A - B|}{\tau_B - \tau_A} \quad . \tag{D1}$$

$\tau_A$ and $\tau_B$ can be defined by the time of flight of the twin photons from the source to the detectors:

$$\tau_A = \frac{|S-A|}{c}, \quad \text{and} \quad \tau_B = \frac{|S-B|}{c}. \tag{D2}$$

Here $c$ is the effective velocity of light along the optical path from S to A, and from S to B. (D1) can be rewritten as,

$$v_{QI} = \frac{|A-B|}{K}c, \tag{11}$$

where, $K = |S-B| - |S-A|$, is the path length difference corresponding to, say, the center of the observed by Bob visibility kink (see, Fig. 2). $v_{QI}$ is then completely defined by the involved here distances, which are our variable experimental parameters. This is an advantage. Note, however, that we need additional assumptions needed to define them, that is to specify the exact meaning of A, B, S in the considered distances [20]. This is, obviously, a drawback, which can be avoided by evaluating $v_{QI}$ as the difference:

$$v_{QI} = \frac{|A_1 - B_1| - |A_2 - B_2|}{K_1 - K_2} c. \tag{12}$$

The new spatial distance, that is the numerator of (12) is well defined experimentally, provided the changes the of positions $A_1 \to A_2$, $B_1 \to B_2$ are parallel shifts of the measuring setups, in which case the involved difference does not depend on the definitions of paths limits. $K_1 - K_2$ is the corresponding new transfer time, which would be better defined experimentally than that in (11), in particular if the shape of the kink is (experimentally) conserved relative to the changes of the distance from A to B.

Now, the experimental evaluation of the lower bound of $v_{QI}$,

$$\min v_{QI} = \frac{|A_1 - B_1| - |A_2 - B_2|}{\delta(K)} c, \tag{13}$$

where $\delta(K)$ is the experimental uncertainty of the $K_1 - K_2$ evaluation, would not depend on the definitions of the involved optical path limits, and would be free of auxiliary assumptions.

The accuracy of the Geneva experiments is not expected to be achieved in the proposed experiment; the direct interest would be in corroborating the former by a different method of measurement, and in getting some new insight into the physics of the quantum information transfer. The latter resides essentially in the two aspects:
(i) The one-way direction of "quantum information transfer" from Alice to Bob.
(ii) The specification of the information content as concerning the decoherence of the system.

This suggests the following conjecture understood as limited to the discussed experiment:

*The decoherence process can be taken for the carrier of the transferred quantum information. The status of the non-local process of decoherence transfer is the same as that of a local decoherence process.*

In other words: *There is no good quantum-mechanical reason to accept a different status of a local decoherence processes, and of a non-local decoherence process* [25].

Note, that the non-local quantum correlation aspects have not been yet explicitly considered in decoherence theory (see, the recent review [24]).

### 5. Test of coherence-conserving measurement

Before applying the proposed basic setup (Fig. 1) to the commutation test of Alice's projective measurement and Bob's interference measurement (see, Sec. 6) it is advisable to have a direct experimental confirmation of the expected coherence-conserving character of Bob's interference measurements. In the case of a positive result, Bob's interference measurement can be regarded as a particular type of 100% non-demolishing measurement in the polarization reduced-Bell-state space. The motive is rather obvious because of the weight of possible consequences. The non-demolishing character of Bob's measurement in the reduced-Bell-state polarization space is the reason behind the expected exception to the standard assumption of the commutation rule of spatially distant measurements [26]. On the other hand, the coherence-conserving test, together with the result of Sec. 3, would complete the experimental confirmation of the main features of the basic setup.

Consider the modified basic setup as shown in Fig. 3. Alice operates now a Mach-Zehnder interferometer equipped with an additional detector, which can be placed in two positions: (0) outside the interferometer, that is non-acting, or (I) in one of the interferometer's arms. Both the Alice's detectors are assumed to act before Bob's detector. Coincidence circuits are linking Alice's and Bob's detectors.

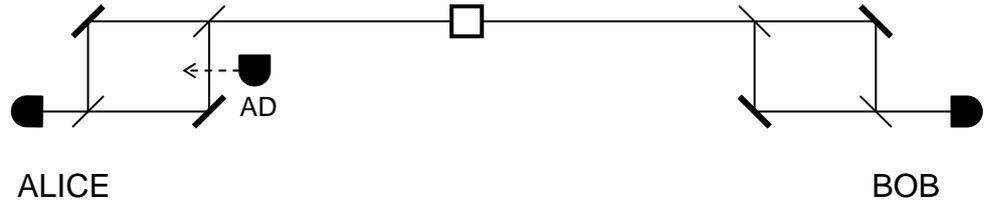

ALICE                                                                 BOB

Fig. 3. The setup to test the non-decoherence of recombined beams measurement. It is a modification of the basic setup of Fig.1: Alice operates the Mach-Zehnder interferometer with the additional detector, AD, which can block one of the partial beams inducing the which-way choice of the B-photon. Both the Alice's detectors are linked to Bob's detector by coincidence circuits. The latter and the details of the basic setup are not shown. Alice's measurements are assumed to precede the interference pattern measurement by Bob.

In case (I) Alice's beam-blocking detector induces the way-choice of the A-photon, and consequently the way-choice of the B-photon resulting in predictions (9) and (10): Bob sees an interference pattern with washed out fringes.

In case (0) Alice's beam-blocking detector does not act, the Alice's interference-fringes-seeing detector records the photons of the recombined beam, avoiding thereby to make the which-way choice. So Bob should see fringes of maximal visibility. But this would mean that Alice's measurement with the recombined beam is coherence-conserving and consequently the main non-demolishing feature of Bob's measurement is occurring in the reduced-Bell polarization state space. The remaining requirements, which should be satisfied by a non-demolition measurement, that is [27]: Bob's measurement distinguishes between coherent and decohered states of the system, Bob's repeated measurements give the same result, are obviously fulfilled.

### 6. Commutation test

The proposed commutation test of Alice's projective measurement and Bob's interference fringes measurement is based essentially on the procedure discussed in Sec. 3. Bob observes the fringes in coincidence with Alice's transmitted beam detector, ATD, as function of the relative detection time. The latter is controlled by varying the optical path length |A- S| and/or |B – S|. Alice's detector in the reflected beam, ARD, is assumed to be non-acting, e.g. removed. The expected fringes visibility, $V_T$, is as shown in Fig. 2. In the next step Bob measures the interference pattern in coincidence with ARD, (visibility, $V_R$) with ATD removed. Both the measurement are expected to yield visibility kinks of the same shape, that is, the expected result is,

$$V_T(|B-S|-|A-S|) \approx V_R(|B-S-|A-S|). \tag{14}$$

Suppose now that Bob measures the fringes, in coincidence with both the ATD and ARD. The resulting visibility, $V_{T+R}$, should be according to (14),

$$V_T \approx V_R \approx V_{T+R}. \tag{15}$$

Here "≈" is understood as the experimental equivalence of the visibility kink as function of the relative measurement time in the A-subsystem and in the B-subsystem. This time difference is experimentally controlled by the involved optical paths length difference (see, Sec. 3). An equivalent $V_{T+R}$ is the outcome of the procedure consisting in the summation of the FD counts from separate coincidence measurements with ATD and ARD. The summation is made at the direct data acquisition level, that is, each point in the observed interference pattern is resulting from the sum of Bob's detector counts with ATD acting and ARD non-acting plus counts with ARD acting and ATD non-acting.

Since the polarization space of the A-subsystem consists of the two (orthogonal) components: the horizontal (transmitted) one, and the vertical (reflected) one, $V_{T+R}$ is resulting from the sum of measurements over the whole polarization space of the A-subsystem, that is from the sum of measurements over all accessible states. Therefore in an experiment in which Bob's detector sees photons from correlated pairs exclusively, here referred to as an ideal experiment, the measured $V_{T+R}$ does not depend on the classical information contribution transmitted by the coincidence circuits. We have to conclude that in an ideal experiment Bob would see $V_{T+R}$ as the outcome of measurements *without the coincidence channels linking Alice's and Bob's detector*. In a real experiment we would expect a less pronounced kink *without coincidences* than the kink *with coincidences*, because of a higher noise level present in the former case. However, since the signal/noise quotient in a real experiment is, roughly, equivalent to the quotient of photons from correlated pairs to the remaining, uncorrelated photons seen by Bob's detector, then according to (2) the number of noise photons in the proposed setup is expected to be relatively low. A real experiment with a noticeable visibility kink without coincidences, that is resulting in outcomes compatible with the inequality (16), should then be feasible.

$$(V_{MAX} - V_{MIN})_{T+R; C(I)} > (V_{MAX} - V_{MIN})_{T+R; C(0)} > 0. \qquad (16)$$

The subscripts "T + R; C(I)" and "T + R; C(0)" label the visibility as measured with coincidences switched on, and with coincidences switched off, respectively.

Let us see once more how the discussed commutation test could run. Assume that the coincidence circuits are switched off. Suppose that in the starting configuration the length of the A-mode optical path is long enough (relative to the B-path length) to let Bob's detector see freely evolving photons. Bob would see then an interference pattern with fringes of maximal visibility, $V_{MAX; C(0)}$ (see, Fig. 4 (a)).

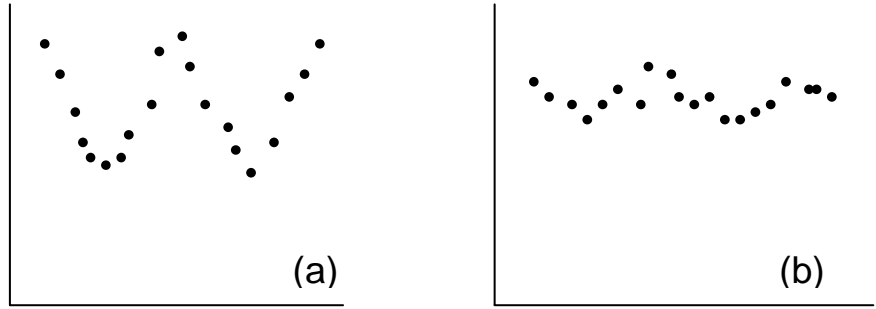

Fig. 4. A qualitative draft of the expected (rough data) interference pattern seen by Bob. (a) Freely evolving system before Alice's projective measurement. (b) System decohered by Alice's measurement.

In the next step (case (b)) the relative length of the optical A-path and of the B-path are reversed, so that of each correlated pair the A-photon makes the way-choice before the B-photon reaches his detector. Bob's detector sees then (apart from the noise photons) the photons, which made the way choice induced by the A-photons choice already. Therefore the fringes in the interference pattern generated by the signal photons are washed out. However, since some of the noise photons with frequency spectrum close to the signal spectrum may form fringes like free signal photons, we may expect an interference pattern as shown in (Fig. 4 (b)). Label the resulting fringe visibility as $V_{MIN; C(0)}$. A positive outcome of the experiment would mean the agreement with (16).

Relation (16) seen (by Bob) without classical channel linking Alice and Bob would be the experimental demonstration of non-commutation of Alice's projective measurement and Bob's interference measurement, since (16) means that Bob's outcome depends on the time order of his and Alice's measurement. Recall that the latter is decohering the superposition in the polarization space of the system, while the former is a polarization-coherence conserving measurement.

### 7. Concluding remarks

The proposed here experimental scheme is based on a hybrid of the polarization-entangled two-photon system, and the double-slit (here a Mach-Zehnder interferometer) setup. The entangled photon pair from spontaneous parametric down-conversion is preselected at the source, so that the polarization plane of both the photons is preset, and then the photons are directed in separate space modes toward the Alice's setup and Bob's setup. Both Alice and Bob split the incoming photons in horizontal and vertical components. Alice makes projective detections in split modes; Bob observes the interference pattern of the recombined beams. The polarization entanglement of the biphoton enables Alice to induce by her projective measurement the way-choice of the Bob's photon, so Alice's remote way-choice washes Bob's interference fringes out. This is the expected first basic result of the proposed scheme (Sec. 2). In the next step the time relation of Alice's projective action and its earliest effect seen by Bob is determined (Sec. 3). This result is referred to as the second basic one. The third basic result would consist in the test of the coherence-conserving character of Bob's interference measurement in the polarization state space of the total system (Sec. 5).

The *first* basic result would provide a new experimental argument in the complementarity-versus-uncertainty-relation issue [13-16], since the which-way-choice would be induced without any direct action on (or of) the interfering photon.

The *second* basic result opens the way to a new evaluation method of the quantum information speed (its lower bound) [17, 20, 21, 23] (Sec. 4). Apart from corroborating the Gisin group results [17, 23], the main interest would be in the new feature: the exclusively one-way (from Alice to Bob) direction of the quantum information transfer, since it is Alice's action which triggers the effect seen by Bob.

The second basic result together with the *third* basic result, which can be interpreted as a 100% non-demolishing measurement in the polarization space, would shed light on the physics of the one-way quantum transfer character from Alice to Bob, and thereby on the physics behind the commutation test of Alice and Bob's measurements (Sec.6). At the *gedanken* level the latter test has to be interpreted as providing an example of non-commutation of Alice and Bob's measurements, although these measurements are mutually separated by a space-like interval.

Thus, we would have to do with *a counterexample to the main assumption of the no-signaling paradigm* [26]. This enables Alice to transfer classical information to Bob by means of the quantum channel exclusively. However, Bob's counting time needed to discriminate experimentally between $V_{MIN; C(0)}$ and $V_{MAX; C(0)}$ makes the classical information transfer extremely slow: roughly, one classical bit per two interference patterns measured. Therefore classical information would be effectively contained in the light cone, even perhaps, in cosmic-scale experiments [28], which makes of the proposed experiment a proof-of-principle one.

The obvious question comes to mind: how seriously should we take the proposed experiment in view of its expected results ?

First of all let me notice that apart from the commutation test (Sec. 6) the proposed experiments are by no means controversial, and their expected results are obviously of interest to foundations issues. Notice further that the expected apparently controversial outcome of the commutation test, that is the non-commutation of Alice and Bob's measurement, has resulted from standard quantum mechanics as applied to the existing experimental knowledge without any auxiliary assumptions.

The tension between quantum mechanics and relativity theory has been a latent foundations problem ever since the appearance of non-locality in the formalism of quantum theory; a problem dwarfed and muffled by the unprecedented success of both the relativity and quantum theory predictions. The "patient merit" of some theoretical arguments (see, for recent examples [29, 30]) has not been often noticed (see, however [31]). On the other hand, the recent experimental access to non-locality via entangled systems generated a revival of the issue.

Since a general discussion is, obviously, beyond the scope of the present paper, I just list some relevant publications[(4)].

---

[(4)] The "peaceful coexistence" of relativity and quantum mechanics: [26, 32]. Beyond quantum mechanics: [33, 34], see, also [23] and refs therein. Preferred frame: [35, 36], see, also [17]. Uncertain coexistence: [37- 42].

---

**Acknowledgments**

I gratefully acknowledge the discussions, encouragement, and help, at different stages of this work, of Robert R. Gałązka, Ludwik Lis, Jakub Rembieliński, Jan Mostowski, Dominik Rogula.